\documentclass[conference]{IEEEtran}
\IEEEoverridecommandlockouts
\usepackage{cite}
\usepackage{amsmath,amssymb,amsfonts}
\usepackage{algorithmic}
\usepackage{graphicx}
\usepackage{textcomp}
\usepackage{xcolor}

\def\BibTeX{{\rm B\kern-.05em{\sc i\kern-.025em b}\kern-.08em
    T\kern-.1667em\lower.7ex\hbox{E}\kern-.125emX}}
\usepackage{blindtext}    
\begin{document}

\title{CNN-DRL for Scalable Actions in Finance\\
{\footnotesize  Short paper (CSCI-RTAI)}
}
 
\author{

\IEEEauthorblockN{1\textsuperscript{nd} Sina Montazeri}
\IEEEauthorblockA{\textit{CSCE Department} \\
\textit{University of North Texas}\\
Denton, TX, USA  \\
SinaMontazeri@my.unt.edu}
\and
\IEEEauthorblockN{2\textsuperscript{rd} Akram Mirzaeinia}
\IEEEauthorblockA{\textit{Independent Researcher} \\
\textit{Not Affiliated }\\
Ajman, United Arab Emirates\\
mirzaei.kad@gmail.com}
\and
\IEEEauthorblockN{3\textsuperscript{rd} Haseebullah Jumakhan}
\IEEEauthorblockA{\textit{Independent Researcher} \\
\textit{Not Affiliated}\\
Ajman, United Arab Emirates \\
hjk608@nyu.edu}
\and
\IEEEauthorblockN{4\textsuperscript{th} Amir Mirzaeinia}
\IEEEauthorblockA{\textit{CSCE Department} \\
\textit{University of North Texas}\\
Denton, TX, USA \\
amir.mirzaeinia@unt.edu}
 }

\maketitle

\begin{abstract}
The published MLP-based DRL in finance has difficulties in learning the dynamics of the environment when the action scale increases. If the buying and selling increase to one thousand shares, the MLP agent will not be able to effectively adapt to the environment. To address this, we designed a CNN agent that concatenates the data from the last ninety days of the daily feature vector to create the CNN input matrix. Our extensive experiments demonstrate that the MLP-based agent experiences a loss corresponding to the initial environment setup, while our designed CNN remains stable, effectively learns the environment, and leads to an increase in rewards.

\end{abstract}


\section{INTRODUCTION}
Reinforcement Learning (RL) is a Machine Learning (ML) algorithm that refines actions through repeated interactions with an environment. The environment provides feedback via states (informative descriptions of the environment) and reward which is the outcomes linked to the prior actions of RL agent and the environment's state. This process enables agents to select actions that maximize cumulative rewards over time. Several distinct Reinforcement Learning (RL) algorithms have been proposed aiming to maximize rewards and minimize costs. Notably, Proximal Policy Optimization (PPO) \cite{schulman2017proximal} and Soft Actor Critic (SAC) \cite{mnih2016asynchronous} are two prevalent RL algorithms known for high performance in diverse applications. We utilize these two algorithms to compare our approach with prior work. 

Applying deep neural networks as function approximators in RL agents enables handling larger, continuous action and state spaces. This combination of deep learning and RL is known as deep reinforcement learning (DRL). While DRL has shown promise for stock trading, a key challenge is effectively handling large, continuous action spaces. As the scale of buy/sell actions increases, instability arises during the training of traditional MLP agents. This stems from the tendency to overfit to narrow ranges of the action space initially explored. Simply adding more nodes to account for larger action scales leads to high dimensionality and sparse gradients.

Our work demonstrates a novel application of convolutional neural networks (CNNs) to address this limitation in financial DRL. The innate ability of CNNs to recognize localized patterns independent of their position makes them well-suited to extract meaningful features despite actions covering a wide span. 

Multi-Layer Perceptron (MLP) is among the initial deep learning models employed in DRL to learn environments with vector state representations, while Convolutional Neural Network (CNN) is commonly used for learning environments with matrix state representations such as in computer vision \cite{krizhevsky2012imagenet}. Research shows CNN models excel in feature extraction within convolutional layers, followed by fully connected layers for classification. Some studies have analyzed DRL approaches for stock trading, such as Pricope's review \cite{pricope2021deep}.  

In this work, we design a CNN architecture as a DRL agent to effectively discern salient features from financial data and achieve high rewards with wide-ranging buy/sell actions. Prior approaches in finrl limit actions to 100 shares, while we increase to 1000 shares.

\section{RELATED WORKS}

A variety of methods have been applied for stock price forecasting and algorithmic trading, including:

\begin{enumerate}
    \item Traditional Statistical Methods: Traditional techniques like autoregressive models (ARIMA) \cite{box2015time}, moving averages, and exponential smoothing \cite{holt2004forecasting} have long seen use in stock price forecasting. However, these methods have limitations in capturing complex nonlinear patterns.

    \item Machine Learning Models: Algorithms like support vector machines (SVM) \cite{kim2003financial}, random forests \cite{ballings2015evaluating}, and gradient boosting \cite{de2018machine} have gained popularity for capturing complex patterns in stock data stock prediction and stock prediction. While powerful, these may not fully utilize the temporal structure of financial data.

    \item Deep Learning Models
    \begin{enumerate}
        \item Recurrent Neural Networks - RNNs and Long Short-Term Memory (LSTM) networks \cite{fischer2018deep} have shown promise by modeling long-term dependencies. However, RNNs can face challenges with vanishing gradients.
        \item 
        Convolutional Neural Networks (CNNs) have been applied by extracting sliding windows of historical price data as input matrices. CNNs can efficiently find localized patterns. Hosseindzade \& Haratizadeh \cite{Hoseinzade_2018} demonstrate the potential of CNN architectures for forecasting price trends and volatility in the financial markets.
    \end{enumerate}
    \item Deep Reinforcement Learning: DRL has become an active area of research for quantitative trading and portfolio management. Some recent papers exploring DRL for stock trading include:
    \begin{enumerate}
         \item Moody et al. \cite{moody1998performance} developed an RRL for intraday trading. It outperformed the Q-learning algorithm in S\&P 500/Tbill data.
        \item Jiang et al. \cite{jiang2017cryptocurrency} applied deep deterministic policy gradient with LSTMs for high-frequency trading. Their model generated stable daily returns.
        \item In NeurIPS 2022 conference Liu et al. published FinRL-Meta \cite{liu2022finrl} which is an openly accessible library that has been actively maintained by the AI4Finance community.
    \end{enumerate}
       
\end{enumerate}

In summary, deep RL shows promise for stock trading but faces challenges in scaling to large action spaces. Our work contributes a tailored CNN architecture to address this limitation.



\section{PROBLEM DESCRIPTION}
Stock trading can be represented as a Markov Decision Process (MDP) with the goal of maximizing the anticipated returns.

\subsection{MDP Model for Stock Trading}
The trading market is a stochastic and interactive environment in nature and can be formulated as a Markov Decision Process (MDP) with state, action, and reward.
\begin{itemize}
    \item State $s=[b, p, h, f]:$ a set that consist of balance $b$, price $p \in \mathbb{R}_+^D$, holdings of stock $h\in \mathbb{Z}_+^D$, and fundamental indicators $f$. where D is the number of stocks that we consider in the market. Fundamental indicators covers financial ratios listed in table \ref{tab:featurelist}.
    \item Action $a=[sell, buy, hold]:$ a set of actions for all D stocks, consisting of sell, buy, hold which leads to a reduction, growth, or no alteration in the holdings h, correspondingly.
    \item Reward $r(s, a, s'):$ The adjustment in portfolio value upon executing action "a" in state $"s"$ and transitioning to the next state $"s'"$. The portfolio value encompasses the total value of equities in the held stocks, denoted as $p^Th$, plus the remaining balance, $"b"$.
    \item Policy $\pi(s)$: The stock trading approach in state "s" entails the probability distribution of "a" in the state "s".
    \item The action-value function $Q_\pi(s, a)$
    represents the anticipated reward obtained by taking action "a" in state "s" according to policy $\pi$.
\end{itemize}
The primary objective of this process is to optimize (maximize) the reward. Various published approaches exist for addressing this challenge, each with its own set of advantages and disadvantages \cite{sutton2018reinforcement}. We select PPO and A2C which are commonly used and show higher performance than other approaches.

\subsection{Environment}
We have updated the environment developed by Finrl to enable the environment state representation in the form of a matrix. Our modified environment initializes the state matrix with data from the initial 90-day period of the dataset. Subsequently, we incrementally shift the matrix by one day each time a new action is taken. There are some monitor/control parameters in the environment to facilitate learning process.

In stock market modeling, turbulence refers to a period of heightened uncertainty, volatility, and instability in the financial markets. It is characterized by rapid and significant fluctuations in stock prices, which can make it difficult to model, predict, and learn to respond to market movements. 
Machine learning models typically exclude the period of uncertainty in the stock market, often characterized by a turbulence factor, in order to stabilize their learning process. In contrast, we chose not to activate the turbulence exclusion and instead closely monitored and took action during this period as well as other period of time.

The Sharpe ratio is the next parameter that measures the risk-adjusted return of an investment or a trading strategy. It was developed by Nobel laureate William F. Sharpe and is widely used in finance to assess the performance of an investment or portfolio relative to its level of risk.
The formula for the Sharpe ratio is:
\begin{equation}
\text{Sharpe Ratio} = \frac{(R_p-R_f)}{\sigma_p}
\end{equation}
Where $R_p$ is the average return of the investment or portfolio. $ R_f$ is the risk-free rate of return (the return of a risk-free investment, typically a government bond).  $\sigma_p$ is the standard deviation of the investment or portfolio's returns, which represents its volatility or risk. 
In this project, this ratio is computed based on the mean divided by the standard deviation of daily assets.

The cost of accumulated reward is another parameter that is closely monitored which corresponds to the number of sell/buy actions that are generated by the DRL agent. In practice, The cost of stock trading refers to the expenses and fees associated with buying and selling stocks in financial markets. These costs can vary depending on various factors, including the type of brokerage or platform used, the frequency of trading, the specific services offered, etc. We employ identical fixed cost percentages as those developed in the FinRL framework. 

\subsection{Feature vector}
 Feature vector encompasses the initial amount, stock price of thirty companies, number of shares held from thirty companies, and fifteen financial ratios for each one of thirty companies as it is listed in table \ref{tab:featurevectore}.

 \begin{table}[htbp]
\caption{Daily Feature Vector}
\begin{center}
\begin{tabular}{|l|l|}
\hline
Name                         & size \\ \hline
Amount                       & 1    \\ \hline
Price                        & 30   \\ \hline
Share held                   & 30   \\ \hline
Financial ratios (15 * 30)     & 450  \\ \hline
Total size of feature vector & 511  \\ \hline
\end{tabular}

\label{tab:featurevectore}
\end{center}
\end{table}

Financial ratios are generally derived from numerical data extracted from a company's financial statements in order to extract significant insights about its performance. The figures presented in financial statements, including the balance sheet, income statement, and cash flow statement, are employed to conduct quantitative analysis, evaluating aspects such as liquidity, leverage, growth, margins, profitability, return rates, valuation, and other relevant metrics.

\begin{table}[htbp]
\caption{Fifteen financial ratios in feature vector}
\begin{center}

\begin{tabular}{|l|l|l|}
\hline
Category & Financial ratios & Metrics                                                                                                                \\ \hline
1        & Liquidity        & \begin{tabular}[c]{@{}l@{}}Current ratio\\ Cash ratio\\ Quick ratio\end{tabular}                                       \\ \hline
2        & Leverage         & \begin{tabular}[c]{@{}l@{}}Debt ratio \\ Debt to equity\end{tabular}                                                   \\ \hline
3        & Efficiency       & \begin{tabular}[c]{@{}l@{}}Inventory turnover ratio\\ Receivables turnover ratio\\ Payable turnover ratio\end{tabular} \\ \hline
4        & Profitability    & \begin{tabular}[c]{@{}l@{}}Operating margin\\ Net profit margin\\ Return on assets\\ Return on equity\end{tabular}     \\ \hline
5        & Market value     & \begin{tabular}[c]{@{}l@{}}Earnings Per Share\\ Book Per Share\\ Dividend Per Share\end{tabular}                       \\ \hline
\end{tabular}

\label{tab:featurelist}
\end{center}
\end{table}



Table \ref{tab:featurelist} lists fifteen financial metrics that are classified into five categories liquidity metrics, leverage indicators, efficiency measures, profitability gauges, and market value assessments. Liquidity ratios assess a company's capacity to settle short- and long-term financial obligations. Leverage ratios quantify the proportion of capital sourced from debt. Essentially, these financial metrics are employed to assess a company's indebtedness. Efficiency ratios, alternatively referred to as activity financial ratios, gauge the effectiveness of a company deploying its assets and resources. Profitability ratios assess a company's capacity to generate earnings in relation to revenue, assets on the balance sheet, operational expenses, and equity. Market value ratios are employed to assess the stock price of a company.

Three types of liquidity ratios are being used including the current ratio, cash ratio, and quick ratio.
The current ratio measures a company’s ability to pay off short-term liabilities with current assets.
The cash ratio measures a company’s ability to pay off short-term liabilities with cash and cash equivalents.
A company's quick ratio is a measure of liquidity used to evaluate its capacity to meet short-term liabilities using its most liquid assets

Two types of leverage financial ratios are used including debt ratio and debt to equity. The debt ratio measures the relative amount of a company’s assets that are provided from debt and the debt-to-equity ratio calculates the weight of total debt and financial liabilities against shareholders’ equity.

Three types of efficiency ratios are being used including inventory turnover ratio, receivables turnover ratio, and payable turnover ratio.
The inventory turnover ratio measures how many times a company’s inventory is sold and replaced over a given period. The accounts receivable turnover ratio measures how many times a company can turn receivables into cash over a given period. the payable turnover is used to quantify the rate at which a company pays off its suppliers.

Four types of profitability ratios are being used including operating margin, net profit margin, return on assets, and return on equity. The operating margin ratio, sometimes known as the return on sales ratio, compares the operating income of a company to its net sales to determine operating efficiency. The net profit margin, or simply net margin, measures how much net income or profit is generated as a percentage of revenue. The return on assets ratio measures how efficiently a company is using its assets to generate profit. The return on equity ratio measures how efficiently a company is using its equity to generate profit

There are three types of valuation ratios that are being used in the feature vector including earnings per share, book per share, and dividend per share. The earnings per share ratio measures the amount of net income earned for each share outstanding. The book value per share ratio calculates the per-share value of a company based on the equity available to shareholders. The dividend yield ratio measures the amount of dividends attributed to shareholders relative to the market value per share.

\section{CNN ARCHITECTURE }
A CNN is a  form of neural network that is essentially designed to work on matrix state representation (2D) data sets such as images and videos. CNN is able to learn features through the optimization of filters (kernels) and it can be formed by a sequence of multiple Convolutional, Pooling, Normalization, Dropout, and Fully Connected layers. The presence of multiple consecutive layers poses training difficulties, primarily attributed to the issues of vanishing and exploding gradients. There are approaches to address the challenges of vanishing and exploding gradients encountered in earlier neural networks by employing regulated weights across fewer connections.

In order to employ 2D Convolutional Neural Networks (CNNs) in financial data analysis, it is imperative to restructure the representation of our environment state into a matrix format, as opposed to a vector. Financial data inherently manifests as a vector composed of daily features, wherein certain attributes like price undergo updates on a daily (or hourly, etc.) basis, while others, including revenue, income, assets, liabilities, inventories, debt, and the like, experience quarterly updates.

\begin{figure}[htbp]
\centerline
{\includegraphics[width=.5\textwidth]{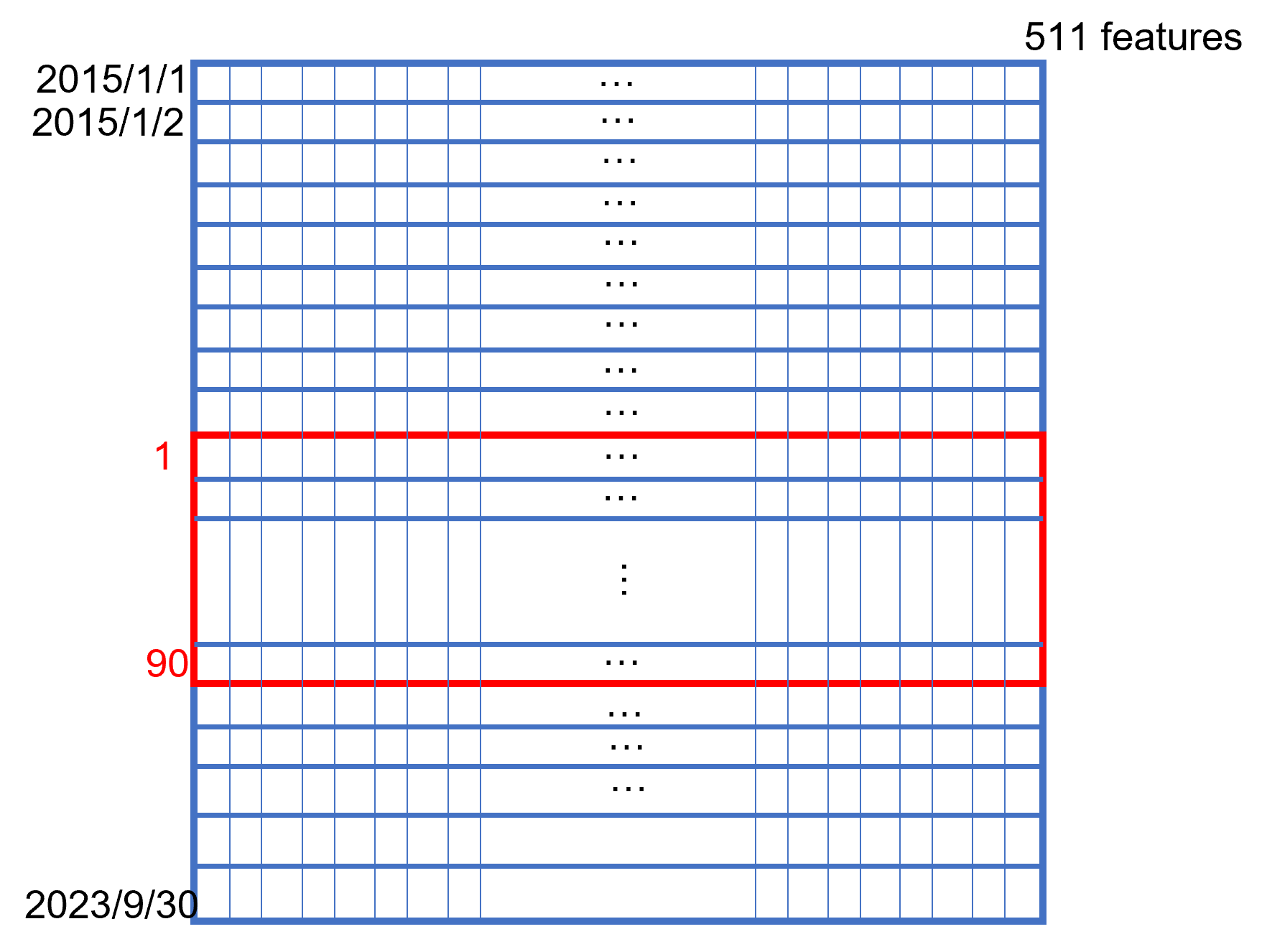}}
\caption{Sliding window to create CNN input matrix channel.}
\label{fig:Sliding_window}
\end{figure}

We designed to concatenate a 90-day feature vector that corresponds to a quarter to create a matrix state representation, Fig. \ref{fig:Sliding_window}. The daily feature vector is the same feature vector published in Finrl \cite{liu2022finrl} which has 511 features extracted for thirty different companies.

\begin{figure}[h]
\centerline
{\includegraphics[width=.5\textwidth]{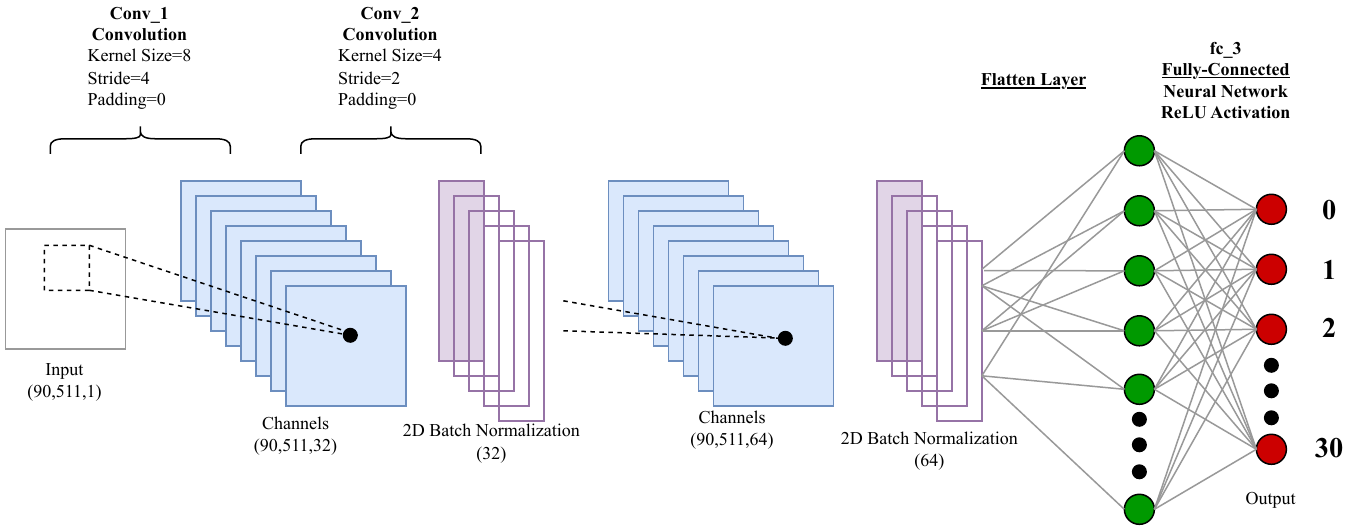}}
\caption{Our CNN architecture.}
\label{fig:CNN_arch}
\end{figure}

Having reconfigured the input into a single-channel sliding window (matrix), some various CNN  architectures have been experimented. It is observed that the CNN architecture faced notable challenges pertaining to vanishing and exploding gradients within this dataset. To address this, we added 2D Batch Normalization layers subsequent to the two convolutional layers Fig. \ref{fig:CNN_arch}. 
This refinement facilitated a notable enhancement in our DRL agent's capability to effectively discern and adapt to the environmental dynamics.

\section{Evaluation}

In order to assess the performance of our designed CNN agent, we employ the FinRL \cite{liu2022finrl}  stock trading environment which is the first open-source framework dedicated to facilitating financial reinforcement learning research. Within the FinRL framework, the daily state of the environment serves as the basis for generating actions that facilitate exploration. These actions manifest as vectors of dimensionality 30, bounded within the interval (-1, +1). Notably, a negative action corresponds to the act of selling a portion of shares, whereas a positive action entails buying some additional shares. To represent the magnitude of the transactions, actions are further scaled to a range of (-1000, +1000), reflecting the number of shares designated for either sale or purchase. This environment utilizes the profile amount as the reward that is initialized to one million dollars. This amount is consistent with prior research. This reward is scaled down to one as it returns from the environment to the agent to help the learning process

We procure the stock prices of thirty companies listed in the Dow Jones from January 2015 through September 2023.  This period encompasses the pandemic, which constitutes one of the challenging times in the realm of finance. This data is fetched from Yahoo Finance and Wharton \cite{Wharton} dataset.

\begin{figure}[h]
\centerline
{\includegraphics[width=.5\textwidth]{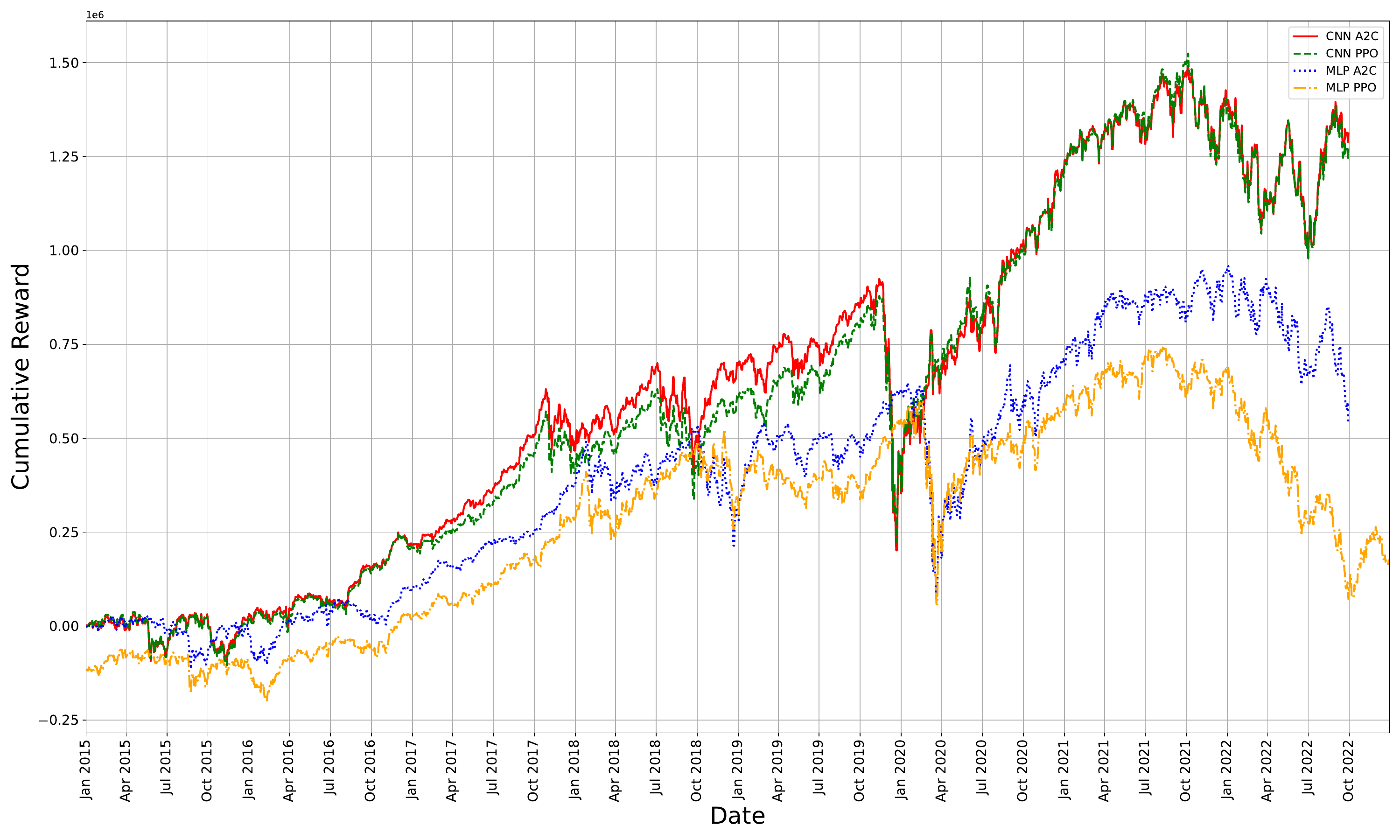}}
\caption{Our CNN vs MLP reward graph}
\label{fig:CNN_MLP_graph}
\end{figure}

Fig. \ref{fig:CNN_MLP_graph} shows the learning reward collected by our CNN and prior MLP agents when two PPO and A2C algorithms are used. As it is demonstrated, our CNN agent is able to learn the environment and increase its reward while the action size is increased to one thousand.

\begin{figure}[h]
\centerline
{\includegraphics[width=.5\textwidth]{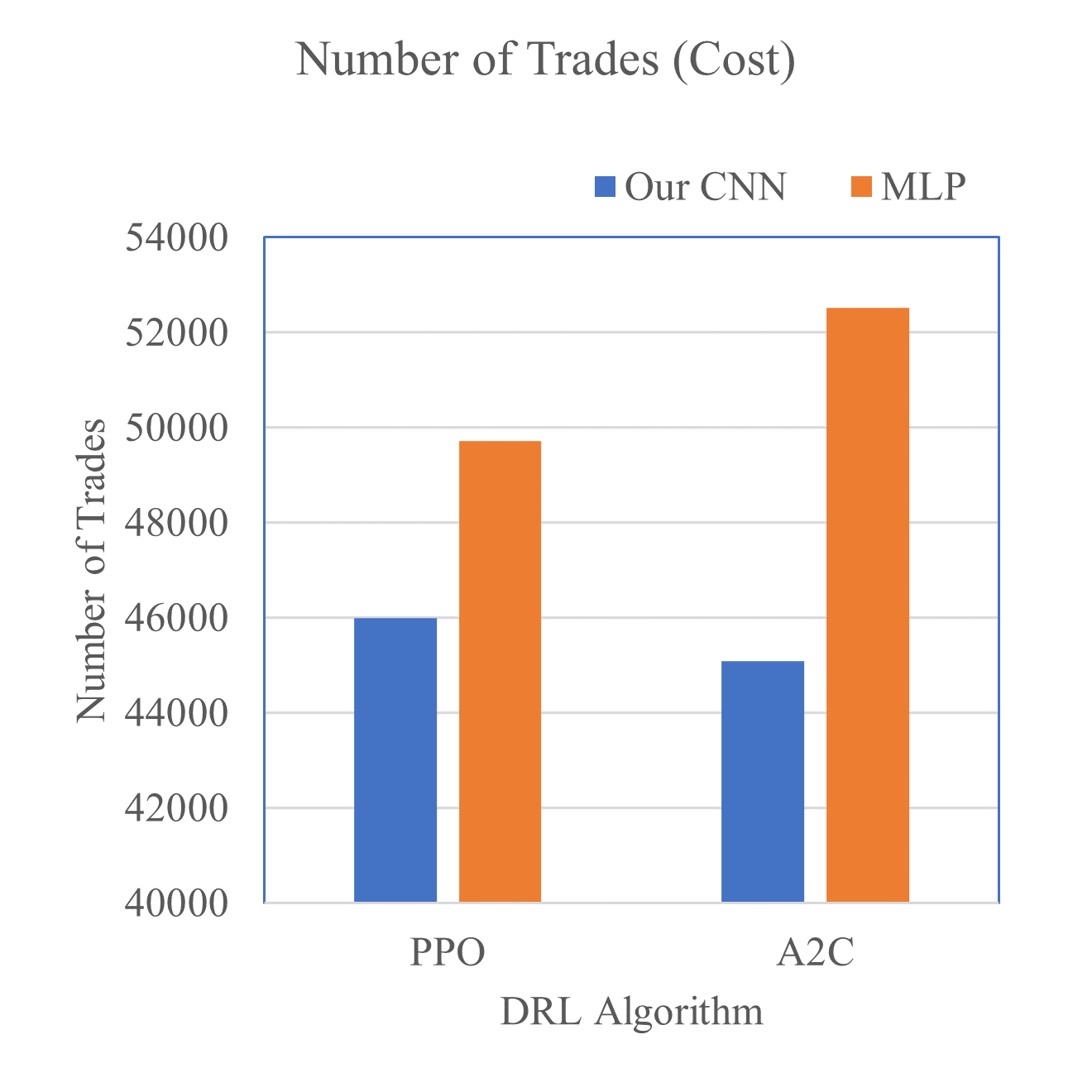}}
\caption{Cost of trading}
\label{fig:train_stat}
\end{figure}

\begin{figure}[h]
\centerline
{\includegraphics[width=.5\textwidth]{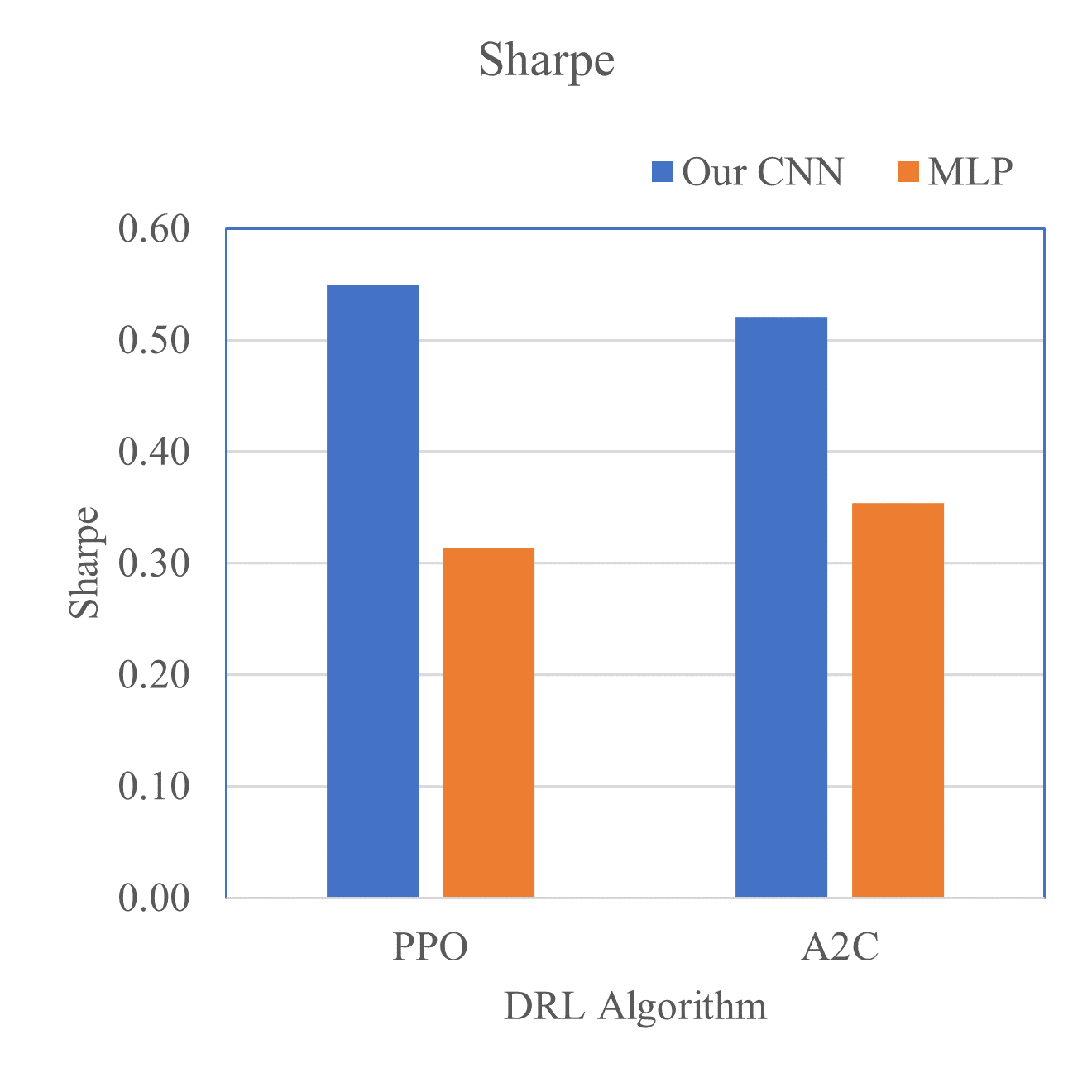}}
\caption{Sharpe}
\label{fig:sharpe}
\end{figure}

Figure \ref{fig:train_stat} presents a comparison of key training statistics between the CNN and MLP agents. As shown, the CNN model achieved better accumulated rewards, while maintaining a lower cost compared to the MLP architecture. Figure \ref{fig:sharpe} highlights the superior risk-adjusted returns (Sharpe ratio) attained by our proposed CNN across both PPO and A2C algorithms.


\section{Conclusion}
This work demonstrates the potential of convolutional neural networks as an effective deep reinforcement learning architecture for stock trading in environments with large, continuous action spaces. Through extensive empirical evaluations using historical market data and the FinRL framework, our tailored CNN agent exhibited stable learning and increased cumulative rewards under expanded buy/sell scales up to 1000 shares. In contrast, traditional MLP models displayed substantial performance degradation when faced with the same enlarged action space.

Our evaluations highlight the inherent aptitude of CNNs for extracting meaningful features from financial data independently of the actions' precise numeric spans. Further research can build on these results by assessing generalizability across other assets, markets, and higher-frequency trading data.

\nocite{*}
\bibliographystyle{IEEEtran} 
\bibliography{biblio}

\vspace{12pt}

\end{document}